# Mission Statement Effect on Research and Innovation Performance

1. Julián David Cortés-Sánchez
    a. Faculty, Universidad del Rosario, Colombia
    b. Invited researcher, FDDI, Fudan University, China. MSc, Development Studies, Universidad de Los Andes, Colombia
2. Jesús María Godoy Bejarano, PhD
    a. Associated professor, Universidad de Ibagué, Colombia
    b. PhD, Management, Universidad de Los Andes, Colombia
3. Diego Téllez, PhD
    a. Assistant professor, Universidad EAFIT, Colombia
    b. PhD, Management, Universidad de Los Andes, Colombia

# Mission Statement Effect on Research and Innovation Performance

**Abstract**

The mission statement (MS) is the most used organizational strategic planning tool worldwide. The relationship between an MS and an organization's financial performance has been shown to be significantly positive, albeit small. However, an MS's relationship to the macroeconomic environment and to organizational innovation has not been investigated. We implemented a Structural Equation Modeling using the SCImago Institutional Ranking (SIR) as a global baseline sample and assessment of organizational research and innovation (RandI), an automated MS content analysis, and the Economic Complexity Index (ECI) as a comprehensive macroeconomic environment measure. We found that the median performance of organizations that do not report an MS is significantly higher than that of reporting organizations, and that a path-dependence driven by the State's long-term view and investment is a better explanatory variable for organizational RandI performance than the MS construct or the intermediate-term macroeconomic environment.

## 1. INTRODUCTION

*"Make it clearer," "add our shareholders," "make it inspiring."* These suggestions could come from low-level employees as well as high-ranking executives when developing a mission statement (MS), the backbone of any organization's strategic plan (Bart et al., 2001; Pearce, 1982). The MS as a strategic planning instrument was originated by scholars and institutions in the United States and gradually spread among scholars and practitioners abroad (Jones, 1960). The content of an MS should express at least four key concepts to the organization's stakeholders: purpose (why the organization exists), values (what it believes in), standards and behaviors (the rules/norms that shape its operations), and strategy (a long-term planning and pathway for achieving its purpose) (Campbell, 1989; Sasaki et al., 2020).

Consider the content of two firms' MS content: Google and Amgen. Google's MS is, "Organize the world's information and make it universally accessible and useful" (Google, n.d.). Amgen's MS is: "To serve patients" (Amgen, n.d.). In terms of readability, a reader would need 16 years of education in American schools to comprehend Google's MS, but only 1.3 years to understand Amgen's MS (Kincaid et al., 1975). While Google's MS does not specifically mention its stakeholders, Amgen's clearly does, namely "patients." The reader can quickly identify which MS is more inspirational and which is more ambitious in tone.

In the end, does internal discussion and debate regarding the content of the company's MS and the costs it represents translate into superior organizational performance? Is superior organizational performance more related to a thriving industry or sector than to the MS content? (Naman and Slevin, 1993; Robinson and Pearce, 1983). Strategic planning activities are not without cost; they represent an average of 25,000 person-days per billion dollars of an organization's revenue (Pfeffer and Sutton, 2006).

These issues have produced a fertile research agenda since the 1980s. In a meta-study summarizing research published from 1987 to 2005, Desmidt et al. (2011) found a significant, albeit small, relationship between MS and financial performance. Still, meta-studies' results have not settled the debate regarding MS and organizational performance. We note four limitations to these meta-studies: (1) sample restrictions, given that the mean sample of organizations analyzed is 137 (Desmidt et al., 2011), (2) narrow focus on certain regions-- most of the studies analyzed organizations located in a single country☐☐and most of them are in developed regions, (3) a focus on financial indicators or financial reporting for measuring organizational performance that excludes other crucial activities such as research and innovation (RandI), and (4) a lack of consideration of the effect of the macroeconomic environment on the content of the MS and on organizational performance.

Despite decades of research, key aspects related to these issues remain unexplored. These gaps lead us to formulate these three research questions:

- RQ 1: do organizations that publicly communicate their MS exhibit higher RandI performance than those that do not?
- RQ 2: does the macroeconomic environment affect MS content?
- RQ 3: is an organization's RandI performance better explained by the content of the MS or the macroeconomic environment?

To answer these RQs, this study examines the effect of the macroeconomic environment on both MS content and RandI performance in multi-sector organizations in a cross-national sample. The sample spans organizations in Africa, Asia, Europe, Latin America and the Caribbean, North America, Oceania, and multinational organizations. These organizations are from governmental, health, higher education, private, and independent sectors. Our sample size ($n$=5,139) is much larger than the sample in Desmidt et al. (2011) ($n$=1,945). This study also considers organizations that do not report their MS on their official website, allowing us to contrast organizational performance between MS reporting and non-reporting organizations. We consider the virtually unexplored issue of performance related to RandI outputs (e.g., research articles output, and scientific and technological impact). It also introduces a factor related to the macroeconomic environment, namely economic complexity. Including the macroeconomic environment could allow us to disentangle RandI's dependence on external factors from an organization's internal ability to plan, design, deploy, and assess its actions.

Furthermore, our MS analysis is comprehensive, analyzing readability features and tone (i.e., lexical diversity and polarity) via automated techniques.

The rest of this study is organized as follows. Section 2 provides a review of the literature on the relationship between MS and performance. Section 3 discusses the methodology used to gather MS, macroeconomic environment variables, content analysis properties, and the proposed Structural Equation Modeling Section 4 presents the results that are discussed in Section 5. Lastly, Section 6 offers our conclusions.

## 2. LITERATURE REVIEW

The interplay between formalized strategic planning, including development of an MS, and organizational performance entails issues such as assessing risk by examining the environment, formulating goals and targets to be achieved, selecting distinctive competencies, determining the relationships between the firms' sub-units, deploying resources to carry out strategies, and controlling and monitoring those activities (Robinson and Pearce, 1983). MS content in particular, and formalized strategic planning inputs in general, must be communicated effectively to maximize their positive organizational impact and support internal leadership, activities, priorities, plans, and work assignments; in sum, to add to excellence in organizational performance (Cochran and David, 1986; Pearce, 1982; Pearce and David, 1987). This alignment summarizes the seminal paths of inquiry of the literature on MS and performance.

Studies on this topic focus on three main areas: comparative performance of firms that report and MS versus those that do not (i.e., non-reporting MS firms); specific characteristics of MS content and MS-orientation; and, the mediating effect of diverse factors (e.g., employees' commitment to the MS) on both financial (i.e., observable and self-reported) and non-financial performance. Most studies analyzed organizations, ranging from private firms to hospitals and universities, from developed countries in their local contexts.

**Comparative Performance on Firms Reporting MS Vs. Non-Reporting MS**

Studies of the comparative performance of organizations with and without an MS have reported mixed results. Bart (1998) found mixed but mostly negative support for the idea that organizations with an MS would perform better than organizations without an MS, based on a sample of large Canadian companies. In contrast, a study of Israeli organizations shows that firms with an MS have better performance than firms with no MS (Sheaffer et al., 2008), while Taghi Alavi and Karami (2009) provide similar results for science parks in the UKand. The latter study emphasized the relationship when either the CEOs/owners or human resources personnel were involved in developing the MS.

**Characteristics of MS Content and MS-Orientation and the Mediating Effect of Diverse Factors**

In a seminal study, Germain and Cooper (1990) argued that firms with a customer service-oriented MS are more likely to survey customers and monitor specific performance metrics (e.g., time per call, number of customers/inquiries, on-time deliveries). Regarding MS-orientation, Atrill et al. (2005) found that among UK firms, MS of service sector organizations that had a shareholders' orientation showed a three-year stock return. In contrast, organizations in the non-service sector that were characterized by an MS with a *stakeholder* orientation, showed a six-year stock return.

Studies focused on firms in developed countries (i.e., the US, Canada, Europe, and Japan) found a positive relationship between specific MS characteristics, namely brevity, a mention of values/beliefs and/or purpose, and no mention of financial goals (Bart and Baetz, 1998), or phrases concerned with *employees*, *social responsibility*, and *values system* (Bartkus et al., 2006) with an increase in the percentage change in sales/profits and return on sales/assets.

Similarly, Jung and Pompper (2014) stated that the MS of the top 50 firms on the Fortune 500 list explicitly mentioned "accommodation" components such as a desired public image, concern for satisfying employees, and/or concern for relationships, and "advocacy" components such as addressing corporate concerns for markets, profits and/or products. Bart et al. (2001) considered the mediating effect of factors such as commitment to the MS, and the degree to which an organization aligns its internal structure, policies and procedures with its MS. Results showed a positive association between employee behavior and the aforementioned factors since employees have the most direct impact on financial performance (i.e., return on sales/assets).

Studies on firms located in developing countries have also focused on the relationship between specific characteristics of MS content and financial performance. MS with RandI-related words were positively associated with superior performance with respect to self-reported and non-financial variables in a sample of Chinese hightech firms (Zhang et al., 2015). In one of the most comprehensive samples reviewed (involving 3,034 Turkish SMEs), Duygulu et al. (2016) stated that three MS components, namely *survival, growth, and profit*; *philosophy and values*; and *public image* were related to firms' performance (i.e., financial, market, production, and overall). In Latin America and Colombia, research has shown a positive relationship between MS readability and financial performance (Cortés-Sánchez and Rivera, 2019) defined in terms of return on assets and return on equity, as well as an effect on asset turnover when an MS promotes good asset management practices (i.e., use of positive language, orientation to financial goals, readability, and asset endowment) (Godoy-Bejarano and Tellez-Falla, 2017).

To our knowledge, the only study analyzing MS content and RandI performance (i.e., patents granted and spin-offs created) found a positive association between particular components or themes of MS, namely customers, product offerings, geographic scope, investors, and society, and performance among organizations in science parks in Spain (Berbegal-Mirabent et al., 2019).

**MS-Performance In the Non-Profit Sector**

In a seminal study conducted in the non-profit sector on Canadian hospitals, Bart and Tabone (1999) affirmed that specific MS components such as distinctive competence/strengths, a unique identity, and a concern for satisfying patients showed a significantly positive correlation with satisfaction with the current MS. Korean hospitals with a diversified MS that identifies target customers or show concern for employees exhibited significantly higher net profits (Kim et al., 2015). Research on not-for-profit health care organizations in Portugal demonstrated that the MS and self-reported performance relationships based on financial measures such as gross income or growth in income, and on non-financial measures such as quality of working environment or donations, are better understood with the mediating effect of organizational commitment (Macedo et al., 2016). Findings on microfinance organizations exhibited a strong coherence between MS focused on alleviating poverty, women's empowerment, and rural financial inclusion, and actual practices (Mersland et al., 2019). Social enterprises that mentioned customers and the product/service offered in their MS were likely to exhibit stronger economic performance (Berbegal-Mirabent et al., 2019).

Evidence from the higher education sector at the institutional level argued that the greater the number of specific sustainability-related terms used in a university's MS (e.g., *sustainable, sustained, sustaining, socially responsible, social responsibility,* etc.) the higher the probability that the university would have a high sustainability rating (Lopez and Martin, 2018). At the academic unit level, Palmer and Short (2008) found in a sample of business schools accredited by the Association to Advance Collegiate Schools of Business (AACSB) that MS were not comprehensive. They found that the MS content was primarily related to business schools' operating budget performance per full-time faculty member or percent of full-time faculty with doctorate degrees. Nonetheless, evidence also shows that the universities' age, focus, and size have a greater explanatory power than MS content characteristics such as readability (Cortés-Sánchez et al., in Press).

MS that seek to build emotional commitment and have semantic attributes such as *activity* were positively associated with key performance metrics (e.g., financial sustainability, donations, number of volunteers) in not-for-profit and performing arts organizations (Pandey et al., 2017; Patel et al., 2015). In sum, MS matter, and 20 years of research on the topic show that MS are positively associated with (financial) performance (Desmidt et al., 2011).

Despite the research reviewed here, there are several gaps in the literature to address, such as the need for a broader and more diversified sample, including macroeconomic factors that could influence both the MS construct and organizational performance, including metrics other than financial indicators and manual MS content analysis, and comparing different geographical regions and economic sectors.

## 3. METHODOLOGY

**Data**

**Organizations' Baseline and RandI Performance**

We used the SCImago Institutions Ranking (SIR) as a baseline database of diverse, research-focused organizations worldwide (SCImago, 2019). The SIR assesses more than 5,000 institutions from the governmental, health, higher education, and private sectors, from over 130 countries, with at least 100 publications indexed in its bibliographic database, Scopus (Bornmann et al., 2012). Institutions are ranked by a composite indicator with a maximum score of 100, which combines three sets of indicators:

- Research performance (output, international collaboration, normalized impact, high-quality publications, excellence, scientific leadership, leadership excellence, scientific talent pool); weight = 50%
- Innovation outputs (number of patent applications, innovative knowledge or publications cited in patents, and technological impact or percentage of scientific publication output cited in patents); weight = 30%
- Societal impact (web sites and inbound links); weight = 20%

Our baseline ranking surpasses the theoretical maximum of 249 ranks stated in recent meta-ranking research based on five organizational innovation worldwide rankings (Lichtenthaler, 2018).

**Mission Statements**

We collected MS manually. For every institution listed in the SIR, a team of five research assistants searched for the institution's MS on its official website between mid-2018 and mid-2019. Since other types of MS-related texts could contain at least one of the eight categories proposed by Pearce (1982), we opted for broader inclusion criteria and searched for text with a similar purpose (e.g., *organization's description; introduction; about us; our purpose; our values; our goals; our objective; what we do,* etc.) even if it was not labeled as an MS. English, Spanish, French, and Portuguese were the languages with the greatest number of MS. The sample analyzed was restricted to MS written in English due to limitations of the automated analysis tools used. Table I presents the summary of English-language texts gathered by region and sector. A total of 215,877 words constituted the MS text corpus.

-------------------------------------------

INSERT TABLE I ABOUT HERE

-------------------------------------------

**Macroeconomic Factors**

Indicators for studying the macroeconomic environment are extensive and diverse, from GPD per capita to the Rule of Law Index. We opted for a single index that significantly explains economic development compared to other indices, the Economic Complexity Index (ECI) (Hidalgo and Hausmann, 2009). Economic complexity "is expressed in the composition of a country's productive output and reflects the structures that emerge to hold and combine knowledge" (for a technical definition, see the Observatory of Economic Complexity (n.d.). The ECI ranks countries based on their export basket diversification and complexity, as countries with a higher diversity in productive know-how are able to produce both a diversified and sophisticated array of products (Hausmann et al., 2014). For instance, a country with a high ECI might produce and export both X-ray machines and microchips; on the other hand, a country with a low ECI might produce and export coffee grains and oil.

Hausmann et al. (2014) compared the ECI with other measures of economic development, such as the six Worldwide Governance Indicators, human capital, and the Global Competitiveness Index. The ECI was shown to be a much more reliable predictor of economic growth than the other indices considered (Hausmann et al., 2014). Furthermore, economic complexity was shown to be a negative and significant predictor of income inequality (i.e., greater complexity is associated with decreasing inequality), and it co-evolves with the inclusivity of a country's institutions (Hartmann et al., 2017). The ECI analyzes more than 120 countries; those with the highest and lowest ECI scores in 2017 were Japan (2.3) and Papua New Guinea (-2.0), respectively (Economic Complexity Observatory, 2017). The complete dataset is available upon request.

**Software and Methods**

We used R (R Core Team, 2014) package Quanteda (Benoit et al., 2018) for content analysis and the AMOS module for SPSS for our SEM (Byrne, 2016; IBM Corp., 2017). Seminal studies (Germain and Cooper, 1990) and recent research ( Berbegal-Mirabent et al., 2019), have used manual content analysis methods. The use of either manual or automated methods depends on the type and length of material to be analyzed (Kondracki et al., 2002). Evidence shows that automated methods for content analysis avoid human biases (Nunez-Mir et al., 2016). However, we opted for a mixed-method since it has been shown to have particularly useful results (Lewis et al., 2013). Specifically, we used manual browsing, capture, and MS classification from each company's website and an automated method for contents analysis.

We propose an MS construct based on three content characteristics: readability, lexical diversity, and polarity. Readability indices examine written materials' comprehensibility with the goal of avoiding unnecessary complexity (Flesch, 1948; Gunning, 1969). Readability indices have been used in management and finance to study annual reports and their effect on analysts' earnings forecasts (Lehavy et al., 2011; Subramanian et al., 1993). Findings have shown that the higher the readability, the higher the performance and understanding of corporate communications (Linsley and Lawrence, 2007; Rennekamp, 2012). Three of the most widely used readability indices are the Gunning Fog index (Cochran, 1986), the Flesch Reading Ease Score, and the Flesch-Kincaid Grade Level Score (Sattari et al., 2011). An analysis of MS of Latin American firms showed a significant correlation among six readability indices, namely the Simple Measure of Gobbledygook, Coleman–Liau, Automated Readability, and the three indices mentioned above (Cortés-Sánchez and Rivera, 2019).

We selected the Flesch-Kincaid Grade Level Score (FKGL). The FKGL was first used to test the readability of technical documents used by the US armed forces (Kincaid et al., 1975). The FKGL calculates the American school grade needed to comprehend a given text. For instance, The Huffington Post's website has an average grade level of about seven, meaning that individuals 12-13 years old should easily understand it. The equation for the FKGL calculation is:

$$FKGL = .39 \left(\frac{words}{sentences}\right) + 11.8 \left(\frac{syllables}{words}\right) - 15.59$$

Source: Kincaid et al. (1975)

As a first attempt to study MS tone, we used the sentiment dimensions proposed by Loughran and McDonald (2011), namely negative, positive, uncertain, litigious, constraining, superfluous, and interesting tones. Preliminary results showed a fairly high lack of such tone characteristics in the MS corpus. A second tone-related analysis focused on two characteristics: lexical diversity and polarity. Lexical diversity refers to the variety of terms in a text (i.e., the number of different words in a given text). Lexical diversity is therefore affected by text size. Tweedie and Baayen (1998) reviewed the most widely used lexical diversity measures to assess their invariance versus text length. They split them into two groups: Yule's K (Yule, 1994) and Z (Tweedie and Baayen, 1998, p. 331). Both indices are theoretically constant in both groups and invariant to text length (Tweedie and Baayen, 1998, p. 249). We chose K as our benchmark for its relative simplicity compared to Z, and for its similarity to models based on *type-token* ratios (TTR). Dillard and Pfau (2002) found that a message receiver prefers information with higher lexical complexity as long as it is within his/her comprehensiveness capacity, reflecting superior credibility and competence from the sender.

Polarity refers to the intensity of the positive or negative feeling in a given text. A lexicon associates each word with a positive or negative feeling. We used the AFINN lexicon (Bradley and Lang, 1999; Nielsen, 2011). It assigns a value ranging from -5, the most negative, to +5, the most positive, to each word. The MS polarity is the algebraic sum of the values assigned to each word. It indicates the positive (polarity>0) or negative (polarity<zero) sentiment conveyed by a given MS. Annex 7.1 shows examples of extreme values of MS in terms of readability, diversity, and polarity. Annexes 7.2 and 7.3 present the number of year-observations for the period 2014-2018, the relative position in the SIR, the descriptive statistics of the ECI, the FKRI, and both polarity and diversity, by regions and quintiles, respectively.

We used SEM to model the effects of ECI on MS construct and on the SIR, and of MS construct on SIR. SEM is consistently used in strategy and management related research (Aguinis et al., 2016; Shook et al., 2004), and offers three main advantages for this study. First, SEM allows one to define latent variables that capture relevant information from a set of observable variables in a model (Byrne, 2016). This allows us to define MS and RandI performance as non-observable constructs and, consequently, to explain the observed variables such as readability for MS, and the SIR for RandI performance. Second, SEM can estimate structural relationships where other methods cannot, given that it allows us to estimate regression coefficients and their significance, departing from co-variance matrices for observable variables (Shook et al., 2004). Third, SEM allows for co-variances between independent variables, a statistical property that captures correlations and interdependencies between independent variables frequently observed in the data (Byrne, 2016).

**4. RESULTS AND ANALYSIS**

The MS is a strategic planning instrument in use worldwide, albeit with regional differences. Sixty percent of the organizations listed in the SIR reported MS-related text on their official websites. Regions such as North America (84%) and Oceania (75%) reported a higher proportion of organizations with public MS-related text, while slightly less than half of Asian (48%) and European (49%) organizations did so.

Table II presents a comparison of the SIR rankings between organizations that report and those that did not report MS-related texts. Results showed that the median rank of non-reporting organizations is significantly higher (closer to 1) than reporting organizations.

--------------------------------------------

INSERT TABLE II ABOUT HERE

--------------------------------------------

Figure 1 shows a scatter plot of a resulting principal component analysis of a random sub-sample of 326 MS (95% confidence; total variation explained by the X-axis = 68.24% and the Y-axis = 10.26%). The top 30 key terms were plotted and segmented into three groups, with size proportional to relative frequency. Comparing this analysis with the six key MS components analyzed by Pearce and David (1982, 1987), the key terms close to the origin (0,0) appear to be related to the principal products/services, market, and technology (e.g., service, technology, teaching, health). The terms in the bottom-right quadrant are related to components of organizations' philosophy, public image, and self-concept (e.g., care, excellence, world, and international). Market- and stakeholder-related components are dispersed (e.g., students, society).

--------------------------------------------

INSERT FIGURE 1 ABOUT HERE

--------------------------------------------

Figure 2 presents the relative frequency of the most common terms for each sector. *Research* is a priority in all except for the private sectors. In the private sector, the top term is *world* as related to reach/influence/incidence or in reference to a timeframe (e.g., tomorrow's *world*) through *innovation* and the development and distribution of *products*. For both the governmental and higher education sectors, *innovation* was not a priority; instead, *knowledge* was most important. It is interesting to note the lack of explicit use of terms related to individuals. Health, higher education, private and other were the only sectors that mentioned terms related to individuals, namely *people, students, community,* and *society,* or *patients*. There were similarities between the health and higher education sectors in terms of references to *education* and *learning. Science* was a priority only for government and other institutions such as the Santa Fe Institute and the RAND Corporation. A plot that analyzed regions did reveal any additional insights.

---------------------------------------------

INSERT FIGURE 2 ABOUT HERE

---------------------------------------------

Figure 3 presents the proposed SEM that estimates the effects of MS and factors related to the economic environment on RandI performance. Both MS and RandI performance are latent variables. MS content is manifested via readability (i.e., FKGL) and tone (i.e., lexical diversity and polarity). Previous studies have used RandI performance measures over one to five years ( Berbegal-Mirabent et al., 2019). We opted to use RandI performance measured by organizations' rankings in the SIR over a three-year period (2016-2018). Other observable variables are innovativeness, mean ECI, and an organization's rank in 2014.

Innovativeness measures the degree of similarity between the sample's MS and the MS corpus of the organizations ranked in the top 5% of SIR in 2014. We proposed an innovativeness variable as an MS benchmark based on the MS of these top performers, as a proxy for the MS content of organizations that have shown superior performance which display particular characteristics (e.g., a *shareholders' orientation,* mentioning values/beliefs, purpose, no mention of financial goals, *employees*, *social responsibility*) (Atrill et al., 2005; Bart and Baetz, 1998; Bartkus et al., 2006). The coincidence level is between zero and one, where one indicates that a given organization MS could be written based on the MS corpus of the top 5% organizations in the SIR. Complexity is the ECI average between 1995-2017 for 133 countries. That allows considering the countries' macroeconomic environment path regarding their progress, stagnation, or decline for 22 years. We also considered the 2014 organizations' rank to integrate into the SIR analysis organizations' persistence between 2014 and 2018.

---------------------------------------------

INSERT FIGURE 3 ABOUT HERE

---------------------------------------------

The content of MS is measured in terms of lexical diversity ($c[M, LD] = 0.984$, $p < 0.001$), readability ($c[M, Re] = 0.366$, $p = NA$), polarity ($c[M, P] = 0.610$, $p < 0.001$) and Innovativeness ($c[M, In] = -0.294$, $p < 0.001$). MS with the highest scores for content variables tend to have a lower coincidence level with the MS corpus of the top 5% organizations as ranked in the SIR. The RandI performance variable is strongly related to the 2016-2018 SIR rankings, but that correlation tends to diminish over time. The organizations' 2014 rankings are highly correlated to their rankings in the period 2016 to 2018; thus, the 2014 ranking emerges as the strongest predictor of RandI performance in subsequent years ($c[R4,IP] = 0.999$, $p < 0.001$), which suggests a strong, persistent effect for the SIR rankings. Factors reflecting the economic environment (i.e., the average ECI over the period 1995-2017), and specifically the effects of economic complexity on MS content ($c[C,M] = 0.037$, $p = 0.008$) and RandI performance ($c[C,IP] = 0.027$, $p < 0.001$) are rather small. The effect of MS on RandI performance ($c[M,IP] = -0.010$, $p = 0.166$) is not significant. This supports the negative relationship between MS and innovativeness ($c[M,In] = -0.294$, $p < 0.001$), given that MS manifests through higher lexical diversity and moderate polarity, but does not fit the crucial innovation-oriented corpus of the MS of the top 5% organizations in the SIR (see Annex 7.4 for more details).

## 5. DISCUSSION

**Exploring the Most Frequent Terms**

Including the term *mission* in the analysis was not an oversight by the authors. It was among the top ten terms in all sectors, and its high relative frequency was especially noticeable in the private sector. This supports organizations' need to express their self-awareness or self-concept, as in *our mission, the university's mission is; the organization's mission; we carry our mission* (Pearce, 1982).

The *Longman Communication 3000* is a list of the 3,000 most frequently used words out of the 390 million that comprise the *Longman Corpus Network* for spoken and written English (Longman Communication 3000, n.d.). Those 3,000 words account for 0.0008% of the total corpus, and for 86% of the entire written English language. Such findings are in line with Zipf's Law (Zipf, 1949). The most common terms, including *research, health, university,* and *world*, are among the top 1,000 most written words in the English language. The two exceptions were *excellence* and *innovation.* A previous content analysis for MS of universities supports these findings (Cortés Sánchez, 2018).

Results support the general orientation of MS in the private sector toward *innovation* (Zhang et al., 2015) and social responsibility (*lives*) (Bartkus et al., 2006). The priority for mentioning *products* is similar for social enterprises and private firms (Berbegal-Mirabent et al., 2019). No sector prioritized mentioning *shareholders*, *employees*, *values*, *survival*, *growth*, or *profits* (Atrill et al., 2005; Duygulu et al., 2016). Explicit mentions of *patients*/*customers* or *employees* in the health sector was not consistent with previous studies (Bart and Tabone, 1999; Kim et al., 2015). Exploration tone characteristics using the sentiment dimensions proposed by Loughran and McDonald (2011) (i.e., negative, positive, interesting, etc.) did not reveal significant results within the MS corpus. Therefore, MS with those tones that are associated with organizational performance in developing countries and in not-for-profit and performing arts entities are exceptions (Godoy-Bejarano and Tellez-Falla, 2017); Pandey et al., 2017); Patel et al., 2015). Based on these results, it appears that open discussions and efforts to make an inspiring, unique, and interesting MS have not produced the desired outcomes.

**Do Organizations That Publicly Communicate Their MS Exhibit Higher RandI Performance Than Those That Do Not?and**

Results showed that the median RandI performance of non-reporting organizations is significantly higher than that of reporting organizations. Explanatory factors could be a function of regional differences in adopting MS, and the existence of research dynamics and ecosystems in regions where the concept of an MS was delayed in its development and implementation.

In the 1960s, Jones (1960) and Levitt (1960) established the seminal pathways for creating the framework and focus for MS composition. Firms then gradually adopted the idea and implemented the MS. To the best of our knowledge, the first empirical study of MS was published in the early 1980s (Pearce, 1982) where the key components of MS were identified. A contemporaneous study that proposed a model for strategic planning in higher education included the MS as a "goal formulation" factor (Kotler and Murphy, 1981). Empirical studies in Europe (specifically, the Netherlands, Germany, and the United Kingdom) emerged in the early-2000s (Cortés Sánchez, 2018; Sidhu, 2003). Even though the Strategic Management Society and the *Strategic Management Journal* are based in London (Nag et al., 2007), 74% of papers published on MS have at least one author affiliated with a United States institution (Scopus, 2018). The time lapse between the initial proposals regarding MS concepts, their practical implementation, and the publication of empirical studies on MS spanned at least 22 years in the United States, and longer in Asian and European countries.

With respect to the sector composition of organizations that do not report an MS, approximately 83% are in higher education and the government sectors (51.7% and 31.1%, respectively). Of the non-reporting organizations in higher education, a total of 77% are in Asia and Europe (44.3% and 32.7%, respectively). Moreover, 91.5% of non-reporting governmental organizations are located in Europe and Asia (69.2% and 22.2%, respectively). Since 80% of the SIR rankings are based on research performance and innovation output, research-intensive organizations and regions in Europe and Asia emphasize research performance assessments regardless of whether or not organizations publicly publish an MS. Evidence consistent with this conclusion includes the following:

- Beijing and Shanghai are among the top five cities of the world's science hotspots (Nature Index, 2020)
- Europe and China have surpassed the US as the world leaders in science and engineering articles published (Tollefson, 2018)
- Over the past 40 years, Europe and Asia have been among the top five regions with the highest number of patents filed (WIPO - the World Intellectual Property Organization, 2019)
- An elite group of public universities of China, known as, the C9 League, (Times Higher Education, 2011) and governmental entities such as the *Max Planck Gesellschaft*, are global research sources

With respect to the sub-sample of private organizations, only 2.3% were in the MS non-reporting group. Ireland and Hitt (1992) identified several reasons (or excuses) companies put forth for not developing an MS, including *no one will read it; too much effort/work for an impractical outcome; is only in an academic exercise; some are successful without one;* and *too much confidential information revealed*, among others. Our findings are consistent with Baetz (1998), who found mixed and mostly negative support for the idea that organizations with an MS should be associated with higher performance compared to organizations without an MS.

**To What Extent Does the Macroeconomic Environment Affect MS Content?**

Our results show the macroeconomic environment has a small effect on MS construction. Previous trans-national research exploring differences between MS, and between MS and performance is sparse, and the evidence suggests no significant differences across MS from organizations in different regions due to differences in macroeconomic environments. MS content analyses for universities worldwide found that MS tend to be longer for universities from Asia and shorter for those in Europe (Cortés Sánchez, 2018). Conversely, significant differences were not found in MS readability among universities from Europe, North America, Asia, Oceania, Latin America, or Africa (Cortés-Sánchez et al., in Press). A study of MS and financial performance in Latin American firms found no discernible differences between MS from Brazil, Chile, or Mexico and their most frequently used keywords, namely clients, products, quality, and shareholders (Cortés-Sánchez and Rivera, 2019).

The idea of isomorphism, i.e., cross-national harmonization due to institutional influences/pressures (Andrews et al., 1993) in MS from universities has been outlined and discussed elsewhere (Cortés Sánchez, 2018). National laws affecting institutes of higher education could reinforce a path toward MS sameness, e.g.*,* "universities should publicly state their commitment to teaching and research" (Kosmützky and Krücken, 2015). In sum, our findings lead to the conclusion that the macroeconomic environment does not significantly affect MS content, with the exception of the MS corpus of the top 5% of organizations in the SIR, considering that it diverges considerably from the MS of the remaining organizations. An explanatory factor with respect to institutional isomorphism in our sample of MS is the harmonization of higher education laws, given that the higher education sector comprise the largest sector of our sample. Among current developments in higher education, Europe is becoming more standardized due to modular and tiered programs prompted by the Bologna process as a mechanism to promote intergovernmental cooperation between 48 countries (European Commission, 2018; Federal Ministry for Education, Science and Culture - Austria, n.d.).

**Is RandI Performance in Organizations Better Explained by MS Content or the Macroeconomic Environment?**

Results showed that MS content and factors related to the macroeconomic environment have a negligible effect on RandI performance in the short run. However, the practical and statistical significance of organizations' 2014 SIR rankings suggests a persistent effect that dampens or overwhelms the short-run impact of MS and contextual factors.

Over 2,120 organizations reported MS and had complete information in SIR for the period 2014-2018. For Asia and Europe, most of these organizations were universities and government organizations, while for the US they were universities, hospitals, and health-related institutions.

Which factors gave those organizations an advantage with respect to their position and persistence? For universities and hospitals/health institutes in the US, basic research productivity has been shifting from corporations to universities since the 1970s (Arora et al., 2019). The National Institutes of Health (NIH), the government's medical research agency, plays a fundamental double-role as a national funding source and research powerhouse. The NIH is made up of 27 institutes and centers, and 156 Nobel Prize winners have received its support (NIH, 2014b, 2014a). The literature on the State's role in promoting radical innovation argues that the NHI had been the nation's and the world's most important investor in medical research since its founding in 1938 (Mazzucato, 2015). The aggregate population of both public (14.5 million) and private colleges (5.1 million) (Statista, n.d., pp. 1965–2028) as a proxy for organizations' size is also a more important explanatory factor regarding US organizations' RandI performance, compared to MS readability (Cortés-Sánchez et al., in Press). In the private sector, corporations such as Amazon, Google/Alphabet, Apple, Tesla, and Netflix that consistently appear in innovation meta analyses, share a common interest in artificial intelligence and a digital platform business model (Lichtenthaler, 2018).

In the short run, the effect of ECI on RandI performance is almost negligible. However, comprehensive studies on the effect of research output on economic growth (i.e., a variable that the ECI explains well) argues that the effect is significantly positive with a few caveats, namely that the effect of research output is exceptionally high in fields such as engineering and technology, social sciences, and physics, and that such effects occurs mainly through structural changes, i.e., by relocating resources toward productive sectors (Pinto and Teixeira, 2020).

Regarding the patenting component of the SIR rankings, various findings have pointed out that half of product innovation is not associated with patenting. Market leaders use patents to deter competitors' innovation and protect their sales (Argente et al., 2020). Conversely, half of product growth and innovation come from firms that do not engage in patenting (Argente et al., 2020). From an historical perspective, attempts to disentangle the effect of innovation on economic growth have argued that growth accelerated after 1750 and reached a maximum in the middle of the 20$^{th}$ century primarily due to two industrial revolutions brought about by the invention of the steam engine, new manufacturing processes and railroads; and by the use of electricity, the internal combustion engine, the development of communications technology, and chemicals (Gordon, 2012). A third industrial revolution (represented by computers, the Internet, and mobile phones) produced a stunted growth between 1996-2004.

# 6. CONCLUSION

Strategic planning processes demand substantial human and financial resources. Among such processes is the formulation of an organization's MS that seeks to highlight the organization's purpose, values, standards and behaviors, and strategy and that must also consider content characteristics such as its readability and uniqueness. A fruitful stream of research has produced a considerable amount of evidence on the relationship between MS and organizational performance from a financial perspective. However, the literature to date has left several questions unaddressed, such as whether organizations that publicly communicate their MS exhibit higher RandI performance than those who do not, whether the macroeconomic environment affects MS content, and whether RandI performance in organizations is better explained by content MS or by the macroeconomic environment. We shed light upon those questions in this study.

The most frequent term used in MS, "exploration," shows the need for private organizations' MS to express self-awareness. We also found extensive use of non-differentiated terms such as *research, health, university,* and *world*. Private sector MS are inclined to emphasize *innovation,* social responsibility (*lives*), and products. *Research* is a priority in all but the private sector. Similarities between the health and higher education sectors were consistent with the themes of *education* and *learning* found in the MS of both. Separately, we find that MS do not contain negative, positive, or interesting tones.

The median RandI performance of organizations that do not report an MS is significantly higher than for those that do report an MS. Since the MS is a strategic planning tool that originated in the United States, both researchers and practitioners in Asia, Europe, LATAM, and Oceania, adopted the practice after it was established in the US. Most non-reporting organizations are in Asia and Europe, regions with large numbers of organizations that produce remarkable research and a substantial number of patents regardless of the lack of a public MS. Reasons for not having an MS, such as the fact that MS development demands too much effort/work for an impractical outcome, were stated in a number of cases in the private sector. The macroeconomic environment showed a small effect on MS due to isomorphism in universities across regions that attempt to harmonize and standardize higher education policies, and to mechanisms seeking to promote intergovernmental cooperation in regions such as Europe. Third, macroeconomic environment factors' effects on organizational performance are negligible due to the complex interactions between the RandI outputs assessed in the SIR and a country's productive output and its structures that are created to create and act as repositories for knowledge, the shift in research activity from the private sector to universities, the decisive role of the government in supporting long-term, high-risk RandI-related activities, and the large size of universities. Among corporations, the persistence of organizations' appearing in innovation rankings is related to a business model that is oriented toward digital platforms and the intensive use of artificial intelligence.

Future studies could trace MS's adoption or modifications in the presence of significant organizational performance change, controlling for environmental and inter-industry factors. Combining various performance indicators, including financial, RandI, and sustainability in an integrated framework, could provide a holistic and integrated understanding of the MS-performance relationship.

**Acknowledgments**

**References**

Aguinis, H., Edwards, J.R., Bradley, K.J., 2016. Improving Our Understanding of Moderation and Mediation in Strategic Management Research: Organizational Research Methods. https://doi.org/10.1177/1094428115627498

Alavi, M.T., Karami, A., 2009. Managers of small and medium enterprises: Mission statement and enhanced organizational performance. Journal of Management Development 28, 555–562. https://doi.org/10.1108/02621710910959729

Amgen, n.d. Mission and Values | Amgen [WWW Document]. Amgen, Inc. URL http://www.amgen.com/about/mission-and-values (accessed 2.28.20).

Argente, D., Baslandze, S., Hanley, D., Moreira, S., 2020. Patents to Products: Product Innovation and Firm Dynamics (No. 2020–4), FRB Atlanta Working Paper, FRB Atlanta Working Paper. Federal Reserve Bank of Atlanta.

Arora, A., Belenzon, S., Patacconi, A., Suh, J., 2019. The changing structure of American Innovation: Some cautionary remarks for economic growth, NBER Working Papers Series. NBER, Massachusetts.

Atrill, P., Omran, M., Pointon, J., 2005. Company mission statements and financial performance. Corporate Ownership and Control 2, 28–35.

Baetz, C.K., Mark C. Bart, 1998. The Relationship Between Mission Statements and Firm Performance: An Exploratory Study. Journal of Management Studies 35, 823–853. https://doi.org/10.1111/1467-6486.00121

Bart, C.K., Baetz, M.C., 1998. The relationship between mission statements and firm performance: An exploratory study. Journal of Management Studies 35, 823–853. https://doi.org/10.1111/1467-6486.00121

Bart, C.K., Bontis, N., Taggar, S., 2001. A model of the impact of mission statements on firm performance. Management Decision 39, 19–35. https://doi.org/10.1108/EUM0000000005404

Bart, C.K., Tabone, J.C., 1999. Mission statement content and hospital performance in the Canadian not- for-profit health care sector. Health Care Management Review 24, 18–29. https://doi.org/10.1097/00004010-199907000-00003

Bartkus, B., Glassman, M., McAfee, R.B., 2006. Mission statement quality and financial performance. European Management Journal 24, 86–94. https://doi.org/10.1016/j.emj.2005.12.010

Benoit, K., Watanabe, K., Wang, H., Nulty, P., Obeng, A., Müller, S., Matsuo, A., 2018. quanteda: An R package for the quantitative analysis of textual data [WWW Document]. Journal of Open Source Software. https://doi.org/10.21105/joss.00774

Berbegal-Mirabent, Jasmina, Alegre, I., Guerrero, A., 2019. Mission statements and performance: An exploratory study of science parks. Long Range Planning 101932. https://doi.org/10.1016/j.lrp.2019.101932

Berbegal-Mirabent, J., Mas-Machuca, M., Guix, P., 2019. Impact of mission statement components on social enterprises' performance. Rev. Manage. Sci. https://doi.org/10.1007/s11846-019-00355-2

Bornmann, L., De Moya Anegón, F., Leydesdorff, L., 2012. The new Excellence Indicator in the World Report of the SCImago Institutions Rankings 2011. Journal of Informetrics 6, 333–335. https://doi.org/10.1016/j.joi.2011.11.006

Byrne, B., 2016. Structural Equation Modeling With AMOS: Basic Concepts, Applications, and Programming, Third Edition. Routledge, Taylor and Francis Group.

Campbell, A., 1989. Does your organisation need a mission? Leadership & Organization Development Journal 10, 3–9. https://doi.org/10.1108/EUM0000000001134

Cochran, D.S., David, F.R., 1986. Communication effectiveness of organizational mission statements. Journal of Applied Communication Research 14, 108–118. https://doi.org/10.1080/00909888609360308

Cortés-Sánchez, J.D., 2018. Mission statements of universities worldwide: Text mining and visualization. Intangible Capital; Vol 14, No 4 (2018)DO - 10.3926/ic.1258.

Cortés-Sánchez, J.D., Bohle, K., Rivera, L., in press. Mission statements in universities: readability and performance. Under review.

Cortés-Sánchez, J.D., Rivera, L., 2019. Mission statements and financial performance in Latin-American firms. 1 20, 270–283. https://doi.org/10.3846/btp.2019.26

Daniel, S.C., 1986. Communication Effectiveness of Organizational Mission Statements. Journal of Applied Communication Research 14, 108–118. https://doi.org/10.1080/00909888609360308

Desmidt, S., Prinzie, A., Decramer, A., 2011. Looking for the value of mission statements: A meta-analysis of 20 years of research. Management Decision 49, 468–483. https://doi.org/10.1108/00251741111120806

Duygulu, E., Ozeren, E., Işildar, P., Appolloni, A., 2016. The Sustainable strategy for small and medium sized enterprises: The relationship between mission statements and performance. Sustainability (Switzerland) 8. https://doi.org/10.3390/su8070698

Economic Complexity Observatory, 2017. OEC - Economic Complexity Ranking of Countries (2013-2017) [WWW Document]. URL https://oec.world/en/rankings/country/eci/ (accessed 11.8.19).

European Commission, 2018. The Bologna Process and the European Higher Education Area [WWW Document]. Education and Training - European Commission. URL https://ec.europa.eu/education/policies/higher-education/bologna-process-and-european-higher-education-area_en (accessed 9.28.20).

Federal Ministry for Education, Science and Culture - Austria, n.d. Current and Future Trends in Higher Education. Wien.

Flesch, R., 1948. A new readability yardstick. Journal of Applied Psychology 32, 221–233. https://doi.org/10.1037/h0057532

Germain, R., Cooper, M.B., 1990. How a customer mission statement affects company performance. Industrial Marketing Management 19, 47–54. https://doi.org/10.1016/0019-8501(90)90027-S

Godoy-Bejarano, J.M., Tellez-Falla, D.F., 2017. Mission Power and Firm Financial Performance. Latin American Business Review 18, 211–226. https://doi.org/10.1080/10978526.2017.1400389

Google, n.d. Google - Information [WWW Document]. URL www.google.com/intl/es/ (accessed 2.28.20).

Gordon, R.J., 2012. Is U.S. Economic Growth Over? Faltering Innovation Confronts the Six Headwinds (Working Paper No. 18315), Working Paper Series. National Bureau of Economic Research. https://doi.org/10.3386/w18315

Gunning, R., 1969. The fog index after twenty years. Journal of Business Communication 6, 3–13. https://doi.org/10.1177/002194366900600202

Hartmann, D., Guevara, M.R., Jara-Figueroa, C., Aristarán, M., Hidalgo, C.A., 2017. Linking Economic Complexity, Institutions, and Income Inequality. World Development 93, 75–93. https://doi.org/10.1016/j.worlddev.2016.12.020

Hausmann, R., Hidalgo, C.A., Bustos, S., Coscia, M., Simoes, A., Yildirim, M.A., 2014. The Atlas of Economic Complexity: Mapping Paths to Prosperity. The MIT Press. https://doi.org/10.7551/mitpress/9647.001.0001

Hidalgo, C.A., Hausmann, R., 2009. The building blocks of economic complexity. Proc Natl Acad Sci USA 106, 10570. https://doi.org/10.1073/pnas.0900943106

IBM Corp., 2017. IBM SPSS Statistics for Windows. IBM Corp., Armonk, NY.

Ireland, R.D., Hitt, M.A., 1992. Mission statements: Importance, challenge, and recommendations for development. Business Horizons 35, 34–42. https://doi.org/10.1016/0007-6813(92)90067-J

Jones, M.H., 1960. Evolving a Business Philosophy. AMJ 3, 93–98. https://doi.org/10.5465/254565

Jung, T., Pompper, D., 2014. Assessing Instrumentality of Mission Statements and Social-Financial Performance Links: Corporate Social Responsibility as Context. International Journal of Strategic Communication 8, 79–99. https://doi.org/10.1080/1553118X.2013.873864

Kim, E.-K., Kim, S.Y., Lee, E., 2015. Analysis of mission statements and organizational performance of hospitals in South Korea. Journal of Korean Academy of Nursing 45, 565–575. https://doi.org/10.4040/jkan.2015.45.4.565

Kincaid, J., Fishburne, R., Rogers, R., Chissom, B., 1975. Derivation of new readability formulas (automated readability index, fog count, and flesch reading ease formula) for Navy enlisted personnel. Chief of Naval Technical Training, Naval Air Station Memphis.

Kondracki, N.L., Wellman, N.S., Amundson, D.R., 2002. Content Analysis: Review of Methods and Their Applications in Nutrition Education. Journal of Nutrition Education and Behavior 34, 224–230. https://doi.org/10.1016/S1499-4046(06)60097-3

Kosmützky, A., Krücken, G., 2015. Sameness and difference: Analyzing institutional and organizational specificities of universities through mission statements. International Studies of Management and Organization 45, 137–149. https://doi.org/10.1080/00208825.2015.1006013

Kotler, P., Murphy, P.E., 1981. Strategic Planning for Higher Education. The Journal of Higher Education 52, 470–489. https://doi.org/10.2307/1981836

Lehavy, R., Li, F., Merkley, K., 2011. The effect of annual report readability on analyst following and the properties of their earnings forecasts. Accounting Review 86, 1087–1115. https://doi.org/10.2308/accr.00000043

Levitt, T., 1960. Marketing Myopia. Harvard Business Review 45–56.

Lewis, S.C., Zamith, R., Hermida, A., 2013. Content Analysis in an Era of Big Data: A Hybrid Approach to Computational and Manual Methods. Journal of Broadcasting & Electronic Media 57, 34–52. https://doi.org/10.1080/08838151.2012.761702

Lichtenthaler, U., 2018. The world's most innovative companies: a meta-ranking. Journal of Strategy and Management 11, 497–511. https://doi.org/10.1108/JSMA-07-2018-0065

Linsley, P.M., Lawrence, M.J., 2007. Risk reporting by the largest UK companies: Readability and lack of obfuscation. Accounting, Auditing and Accountability Journal 20, 620–627. https://doi.org/10.1108/09513570710762601

Longman Communication 3000, n.d. Longman Communication 3000 [WWW Document]. Longman Communication 3000. URL https://www.lextutor.ca/freq/lists_download/longman_3000_list.pdf

Lopez, Y.P., Martin, W.F., 2018. University Mission Statements and Sustainability Performance. Business and Society Review 123, 341–368. https://doi.org/10.1111/basr.12144

Loughran, T., McDonald, B., 2011. When Is a Liability Not a Liability? Textual Analysis, Dictionaries, and 10-Ks. The Journal of Finance 66, 35–65. https://doi.org/10.1111/j.1540-6261.2010.01625.x

Macedo, I.M., Pinho, J.C., Silva, A.M., 2016. Revisiting the link between mission statements and organizational performance in the non-profit sector: The mediating effect of organizational commitment. European Management Journal 34, 36–46. https://doi.org/10.1016/j.emj.2015.10.003

Mazzucato, M., 2015. The entrepreneurial state: debunking public vs. private sector myths. PublicAffairs, US.

Mersland, R., Nyarko, S.A., Szafarz, A., 2019. Do social enterprises walk the talk? Assessing microfinance performances with mission statements. Journal of Business Venturing Insights 11. https://doi.org/10.1016/j.jbvi.2019.e00117

Nag, R., Hambrick, D.C., Chen, M.-J., 2007. What is strategic management, really? Inductive derivation of a consensus definition of the field. Strategic Management Journal 28, 935–955. https://doi.org/10.1002/smj.615

Naman, J.L., Slevin, D.P., 1993. Entrepreneurship and the concept of fit: A model and empirical tests. Strategic Management Journal 14, 137–153. https://doi.org/10.1002/smj.4250140205

Nature Index, 2020. Nature Index's top five science cities, by the numbers [WWW Document]. URL https://www.natureindex.com/news-blog/nature-index-top-five-science-cities-research-by-the-numbers?fbclid=IwAR1CVu0kBIRqIQHbyn6kEnAHoqf7LmujVQT2HouCU8Kzh5Ts68tkbbir56Q (accessed 9.28.20).

NIH, 2014a. Organization - NIH [WWW Document]. National Institutes of Health (NIH). URL https://www.nih.gov/about-nih/who-we-are/organization (accessed 8.13.20).

NIH, 2014b. History - NIH [WWW Document]. National Institutes of Health (NIH). URL https://www.nih.gov/about-nih/who-we-are/history (accessed 8.13.20).

Nunez-Mir, G.C., Iannone, B.V., Pijanowski, B.C., Kong, N., Fei, S., 2016. Automated content analysis: addressing the big literature challenge in ecology and evolution. Methods in Ecology and Evolution 7, 1262–1272. https://doi.org/10.1111/2041-210X.12602

Observatory of Economic Complexity, n.d. Methodology [WWW Document]. URL https://oec.world/en/resources/methodology/ (accessed 11.8.19).

Palmer, T.B., Short, J.C., 2008. Mission statements in U.S. colleges of business: An empirical examination of their content with linkages to configurations and performance. Academy of Management Learning and Education 7, 454–470.

Pandey, S., Kim, M., Pandey, S.K., 2017. Do Mission Statements Matter for Nonprofit Performance?: Insights from a Study of US Performing Arts Organizations. Nonprofit Management and Leadership 27, 389–410. https://doi.org/10.1002/nml.21257

Patel, B.S., Booker, L.D., Ramos, H.M., Bart, C., 2015. Mission statements and performance in non-profit organisations. Corporate Governance (Bingley) 15, 759–774. https://doi.org/10.1108/CG-07-2015-0098

Pearce, J.A., 1982. The company mission as a strategic tool. Sloan Management Review 23, 15.

Pearce, J.A., David, F., 1987. Corporate Mission Statements: The Bottom Line. Academy of Management Executive (08963789) 1, 109–115.

Pfeffer, J., Sutton, R., 2006. Hard facts, dangerous half-truths & total nonsense: profiting from evidence-based management. Harvard Business School Publishing, Boston: United States.

Pinto, T., Teixeira, A.A.C., 2020. The impact of research output on economic growth by fields of science: a dynamic panel data analysis, 1980–2016. Scientometrics 123, 945–978. https://doi.org/10.1007/s11192-020-03419-3

Powell, W., Dimaggio, P., 1993. The New Institutionalism in Organizational Analysis. Administrative Science Quarterly 38, 691.

R Core Team, 2014. R: A language and environment for statistical computing. R Foundation for Statistical Computing, Vienna, Austria.

Rennekamp, K., 2012. Processing Fluency and Investors' Reactions to Disclosure Readability. Journal of Accounting Research 50, 1319–1354. https://doi.org/10.1111/j.1475-679X.2012.00460.x

Robinson, R.B., JR., Pearce, J.A., II, 1983. The impact of formalized strategic planning on financial performance in small organizations. Strategic Management Journal 4, 197–207. https://doi.org/10.1002/smj.4250040302

Sasaki, I., Kotlar, J., Ravasi, D., Vaara, E., 2020. Dealing with revered past: Historical identity statements and strategic change in Japanese family firms. Strategic Management Journal 41, 590–623. https://doi.org/10.1002/smj.3065

Sattari, S., Pitt, L.F., Caruana, A., 2011. How readable are mission statements? An exploratory study. Corporate Communications: An International Journal 16, 282–292. https://doi.org/10.1108/13563281111186931

SCImago, 2019. Scimago Institutions Rankings [WWW Document]. Scimago Institutions Rankings. URL https://www.scimagoir.com/ (accessed 9.10.19).

Scopus, 2018. Search [WWW Document]. URL https://www.scopus.com/

Sheaffer, Z., Landau, D., Drori, I., 2008. Mission statement and performance: An evidence of "Coming of Age." Organization Development Journal 26, 49–62.

Shook, C.L., Ketchen, D.J., Hult, G.T.M., Kacmar, K.M., 2004. An assessment of the use of structural equation modeling in strategic management research. Strategic Management Journal 25, 397–404. https://doi.org/10.1002/smj.385

Sidhu, J., 2003. Mission Statements:: Is it Time to Shelve Them? European Management Journal 21, 439–446. https://doi.org/10.1016/S0263-2373(03)00072-0

Sinclair, S., Geoffrey, R., 2012. About Voyant Tools | Voyant Tools Documentation. URL http://digihum.mcgill.ca/voyant/about/ (accessed 8.3.19).

Statista, n.d. U.S. college enrollment statistics 1965-2028 [WWW Document]. Statista. URL https://www.statista.com/statistics/183995/us-college-enrollment-and-projections-in-public-and-private-institutions/ (accessed 8.13.20).

Subramanian, R., Insley, R.G., Blackwell, R.D., 1993. Performance and Readability: A Comparison of Annual Reports of Profitable and Unprofitable Corporations. Journal of Business Communication 30, 49–61. https://doi.org/10.1177/002194369303000103

Times Higher Education, 2011. Eastern stars: Universities of China's C9 League excel in select fields [WWW Document]. URL https://www.timeshighereducation.com/news/eastern-stars-universities-of-chinas-c9-league-excel-in-select-fields/415193.article (accessed 6.18.20).

Tollefson, J., 2018. China declared world's largest producer of scientific articles. Nature 553, 390–390. https://doi.org/10.1038/d41586-018-00927-4

WIPO - World Intellectual Property Organization, 2019. Patent Cooperation Treaty Yearly Review – 2019. Geneva.

Zhang, H., Garrett, T., Liang, X., 2015. The effects of innovation-oriented mission statements on innovation performance and non-financial business performance. Asian Journal of Technology Innovation 23, 157–171. https://doi.org/10.1080/19761597.2015.1074495

Zipf, G.K., 1949. Human behavior and the principle of least effort., Human behavior and the principle of least effort. Addison-Wesley Press, Oxford, England.

# ANNEXES

# Examples of MS by Readability and Tone Characteristics.

**Table I Examples of MS by readability and tone characteristics**

| Characteristic | Index | Organization | MS |
|---|---|---|---|
| Readability | 1.3 | Amgen | To Serve Patients |
| | 108.7 | Rashtrasant Tukadoji Maharaj Nagpur University | To provide greater access for higher education to all and in particular to the socially and educationally unprivileged youth upholding the principle of social equity; to promote academic excellence and innovation through state-of-the-art Undergraduate, Postgraduate, Doctoral and Post-Doctoral programs; to make the education globally competitive and socioeconomically relevant through competent faculty, modern infrastructure and technology support; to enhance the status of Departments, Conducted and Affiliated Colleges in the fields of knowledge generation and dissemination by pro-actively supporting cutting edge research; to offer educational programs catering the current and future needs of society, region and industries; to provide inspiring conducive academic, social and cultural experience atmosphere to the students, teachers and staff facilitating realization of their full potential and all round development; to increase the efficiency, transparency and accountability in govenance upholding the best interest of the students and the community; to reform examination system for improving standard of evaluation and weeding out systemic inefficiencies; to ensure continued adequate funding through knowledge enterprise and efficient resource management; to strive for community welfare through extension services involving youth in order to cultivate the spirit of integration and sense of ownership as well as social responsibility in them; to acquire and nurture creative human resources and upgrade excellence and skills of existing staff; extensive use of ICT for teaching, information dissemination, administrative, financial, examination processes and transactions between students, colleges and various departments of university; to encourage and facilitate inter-institutional and international exchange programs and collaborations in teaching and research. |
| Lexical diversity | 1.414 | Kuopio University Hospital | To promote health |
| | 15.92 | Colgate University | Colgate's mission is to provide a demanding, expansive, educational experience to a select group of diverse, talented, intellectually sophisticated students who are capable of challenging themselves, their peers, and their teachers in a setting that brings together living and learning. The purpose of the university is to develop wise, thoughtful, critical thinkers and perceptive leaders by challenging young men and women to fulfill their potential through residence in a community that values intellectual rigor and respects the complexity of human understanding […] |
| Polarity | -26 | Institut de Diagnosi Ambiental i Estudis de l'Aigua | It is difficult to overemphasize the potential value of high-quality, hard science based research disseminated on an international stage in the fight to help resolve, or at least ameliorate, 21st century environmental problems as extreme as hydrologic sustainability, megacity air quality, and the ongoing global extinction event affecting so many species in our ecosystems. Human society is expanding and globalizing at an unprecedented, accelerating rate, pushing the limits |

| | | | of what our ecosystems can sustain. Increased water use and severe scarcity, especially in arid and semiarid regions, have been highlighted by the World Economic Forum as a global risk. Water shortages result not from the global lack of water, but from the spatial and temporal mismatch between demand and supply, and things are going to worsen as our collective demand increases. Water engineers have traditionally overcome the problem using reservoirs, water transfers, desalination and groundwater, but all these sources and solutions are by now severely stressed, and the water is increasingly polluted with a wide range of emerging contaminants. Of particular global concern are emerging micropollutants (EMP's) and persistent organic pollutants (POPs) which can biomagnify and bioaccumulate in ecosystems, inducing toxic effects that can be as poorly understood as they are potentially pernicious. In the field of air quality, we are similarly facing an environmental problem of unprecedented scale. Realistic calculations by the UN predict 5 billion people being added to current urban populations by 2050, with nearly 90% of the increase concentrated in Asia and in Africa. Megacities reaching populations of 100 million people are envisaged, creating environmental challenges that are hard to imagine. New approaches are required. In the case of water supplies, Integrated Water Resources Management Groundwater is emerging as the way to address water scarcity, combining surface and ground water and pristine and wastewater to reconcile the demands of people, agriculture, industry and the environment. Society is recognizing that it no longer has the luxury of using water only once. With regard to water cleanliness, the current trend is to reduce the production of residues and the use of chemical products during depuration processes, leaning more toward the natural treatment for supply water production. Finally, all of us living in cities are increasingly aware of the fact that we breathe air contaminated with toxic particles and gases. In Europe, the main problem is road traffic, but in the developing world these combine with industrial and domestic emissions and poor infrastructure to create what is a global insult to human health and results in millions of premature deaths. We need to develop more efficient and accurate methods of measuring our daily dosage of these pollutants, draw up legislative controls that really make a difference, tell people exactly what they are breathing and why, and find new solutions such as hightech purifiers and low-emission transport in greener cities where clean air is a priority demanded by its citizens. |
| | 68 | Colgate University | [Shown above.] |

Source: the authors' based on organizations' websites.

## Descriptive Statistics of the SRI, the ECI, Polarity, and Diversity by Region

**Table II. Descriptive statistics of the SRI, the ECI, polarity, and diversity by region**

| Region | Obs. | SRI Median (IQR) | ECI Mean (s.d) | FKRI Mean (s.d) | Polarity* Mean (s.d) | Lexical Diversity* Mean (s.d) |
|---|---|---|---|---|---|---|
| Africa | 425 | 649 (59.0) | -0.50 (0.70) | 22.1 (6.2) | 7.0 (6.3) | 5.3 (1.6) |
| Asia | 3745 | 603 (112.0) | 1.01 (0.90) | 20.4 (8.3) | 6.7 (6.3) | 5.4 (1.5) |
| Europe | 3715 | 557 (178.0) | 1.32 (0.54) | 20.0 (9.2) | 6.0 (6.4) | 5.5 (2.0) |
| LATAM-CAR | 85 | 629 (119.2) | 0.21 (0.13) | 23.0 (8.4) | 2.3 (2.2) | 5.0 (1.7) |
| MUL (Multinational institutions) | 155 | 378 (247.0) | | 13.8 (4.6) | | |
| North America | 3505 | 531 (247.0) | 1.48 (0.28) | 18.9 (6.9) | 8.0 (7.6) | 5.2 (1.9) |
| Oceania | 415 | 538 (129.0) | -0.27 (0.23) | 16.6 (6.8) | 4.8 (4.5) | 4.3 (1.3) |
| All | 12045 | 568 (169.0) | 1.15 (0.77) | 19.7 (8.2) | 6.9 (6.8) | 5.3 (1.8) |

Source: the authors' based on SCImago (2019); Economic Complexity Observatory (2017); and organizations' websites. Note: * statistics computed after removing missing values and outliers.

**Descriptive Statistics of the SRI, the ECI, Polarity, and Diversity by Quintiles**

**Table III Descriptive statistics of the SRI and the ECI by quintiles**

| Quintile | Obs. | SIR Median (IQR) | ECI Mean (s.d) | FKRI Mean (s.d) | Polarity* Mean (s.d) | Lexical Diversity* Mean (s.d) |
|---|---|---|---|---|---|---|
| 1 | 2680 | 299 (218) | 1.5 (0.5) | 18.5 (6.9) | 5.5 (6.77) | 5.2 (1.7) |
| 2 | 2345 | 512 (59) | 1.3 (0.7) | 18.5 (7.0) | 6.3 (5.6) | 4.9 (1.6) |
| 3 | 2330 | 573 (40) | 1.2 (0.7) | 18.7 (7.5) | 6.5 (5.6) | 5.1 (1.7) |
| 4 | 2510 | 624 (35) | 1.0 (0.8) | 21.1 (9.0) | 7.2 (6.8) | 5.5 (1.7) |
| 5 | 2180 | 661 (34) | 0.6 (0.9) | 22.2 (9.6) | 8.8 (9.0) | 5.9 (2.0) |
| All | 12045 | 568 (169) | 1.2 (0.8) | 19.7 (8.2) | 6.9 (6.8) | 5.3 (1.8) |

Source: the authors' work, based on SCImago (2019); Economic Complexity Observatory (2017); and organizations' websites. Note: *statistics computed after removing missing values and outliers.

## MLE Estimates of the Structural Equation Model

**Table IV MLE estimates of the Structural Equation Modeling**

| Variables (Explanatory ---> Explained) | | Estimate | Std. error. | C.R. | P values | Standardized estimates |
|---|---|---|---|---|---|---|
| **Measurement model** | | | | | | |
| Mission | ---> Readability | 1.000 | | | 0.938 | |
| Mission | ---> Polarity | 13.066 | 1.211 | 10.789 | *** | 0.610 |
| Mission | ---> Lexical Diversity | 5.523 | .430 | 12.839 | *** | 0.984 |
| Mission | ---> Innovativeness | -.096 | .012 | -7.920 | *** | -0.294 |
| Innovation Performance | ---> Ranking (2017) | .924 | .009 | 104.494 | *** | 0.972 |
| Innovation Performance | ---> Ranking (2018) | 1.000 | | | | 0.938 |
| Innovation Performance | ---> Ranking (2016) | .945 | .012 | 77.118 | *** | 0.979 |
| **Structural model** | | | | | | |
| Mission | ---> Innovation Performance | -4.168 | 3.011 | -1.384 | .166 | -0.10 |
| **Regressions** | | | | | | |
| Complexity | ---> Mission | .015 | .006 | 2.666 | .008 | 0.037 |
| Ranking (2014) | ---> Mission | .000 | .000 | 4.565 | *** | 0.148 |
| Complexity | ---> Innovation Performance | 4.674 | 1.250 | 3.740 | *** | 0.027 |
| Ranking (2014) | ---> Innovation Performance | 1.036 | .013 | 80.594 | *** | 0.999 |

| **Variances** | | | | | |
|---|---|---|---|---|---|
| Complexity | .620 | .026 | 23.467 | *** | |
| Ranking (2014) | 17379.600 | 740.589 | 23.467 | *** | |
| e_Mission | .099 | .016 | 6.206 | *** | |
| e_Innovation Performance | 415.920 | 59.417 | 7.000 | *** | |
| e_Lexical diversity | .100 | | | | |
| e_Readability | .657 | .028 | 23.351 | *** | |
| e_Polarity | 29.162 | 1.266 | 23.027 | *** | |
| e_Ranking 2017 | 922.835 | 58.932 | 15.659 | *** | |
| e_Ranking 2018 | 2571.100 | 123.369 | 20.841 | *** | |
| e_Ranking 2016 | 729.092 | 59.378 | 12.279 | *** | |
| e_Innovativeness | .010 | .000 | 23.398 | *** | |

Notes: e_variable refers to the unexplained variance of variable. *** refers to coefficients with p-value below 0.001. The fit statistics are Chi-sq = 1062.52 (285 df), p-value = 0.000, GFI = 0.961, AGFI = 0.919, CFI = 0.984, TLI = 0.974, NFI = 0.979, RMSEA = 0.022.

# TABLES AND FIGURES

## Tables

**Table V Text gathered by region and sector**

| Region | Sector | | | | | |
|---|---|---|---|---|---|---|
| | Government | Health | Higher Education | Private | Others | **Total** |
| Africa | 7 | 7 | 82 | | | 96 |
| Asia | 128 | 68 | 589 | 31 | 2 | 818 |
| Europe | 253 | 207 | 447 | 24 | 7 | 938 |
| Europe-Asia | 2 | | 32 | 1 | | 35 |
| LATAM-CAR | 33 | 21 | 136 | | | 190 |
| North America | 62 | 186 | 406 | 34 | 22 | 710 |
| Oceania | 6 | 34 | 32 | 1 | 3 | 76 |
| MUL (Multinational institutions) | | | | 1 | 15 | 16 |
| Total | 491 | 523 | 1725 | 106 | 34 | 2879 |

Source: the authors' work, based on the organizations' website. Note: Multinational institutions (MUL) are organizations that cannot be attributed to any country. LATAM-CAR is Latin America and the Caribbean.

**Table VI Comparison of rankings between reporting and non-reporting organizations of mission statements**

| Year | Ranking (Median) | | Wilcoxon Test | | F Test | |
|---|---|---|---|---|---|---|
| | Reporting | Non-reporting | W | P-Value | F stat. | P-value |
| **2014** | 603 | 560 | 1.97E6 | <0.001 | 143.53 | <0.001 |
| **2015** | 602 | 561 | 2.08E6 | <0.001 | 140.15 | <0.001 |
| **2016** | 612 | 572 | 2.16E6 | <0.001 | 131.44 | <0.001 |
| **2017** | 609 | 565 | 2.55E6 | <0.001 | 130.47 | <0.001 |
| **2018** | 655 | 607 | 2.41E6 | <0.001 | 122.53 | <0.001 |

Notes: Comparison of rankings between organizations reporting an MS (*n*=2,879) and non-reporting organizations (*n*=2,260). Headers refer to the Wilcoxon rank-sum test (non-parametric test) with continuity correction and one-way test (Parametric test). Ranking refers to the relative position in the SCImago ranking (1=highest). Source: the authors' work, based on SCImago (2019) and organizations' websites.

**Figures**

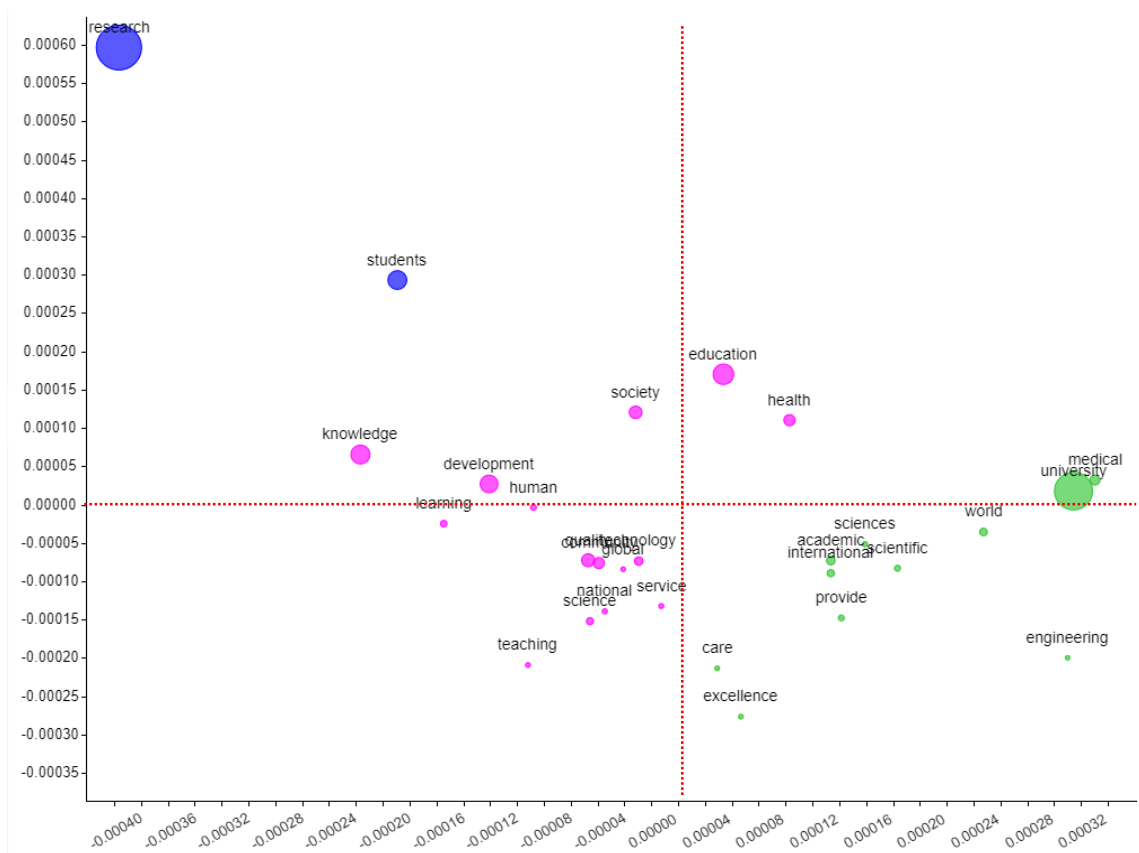

**Figure 1 Scatter plot of principal component analysis of a sub-sample of the MS. Top-30 key terms are shown. Source: the authors' work based on SCImago (2019) and organizations' websites, processed with Voyant-Tools (Sinclair and Geoffrey, 2012).**

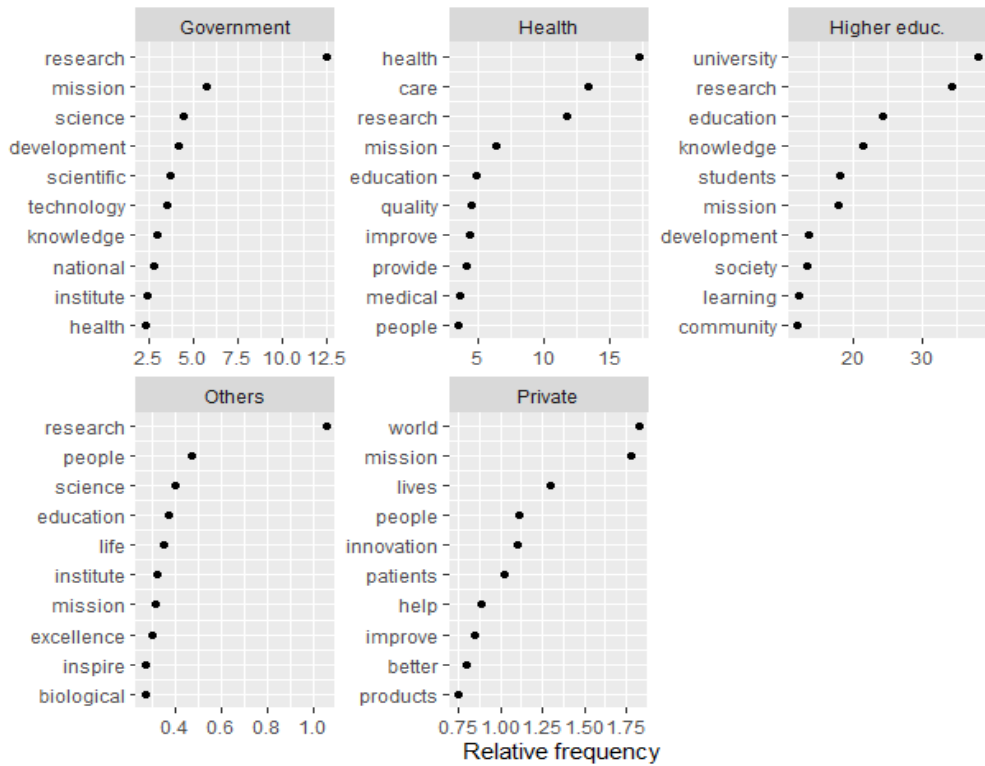

**Figure 2** Top-ten most frequent terms by sector. Source: the authors' work based on organizations' websites.

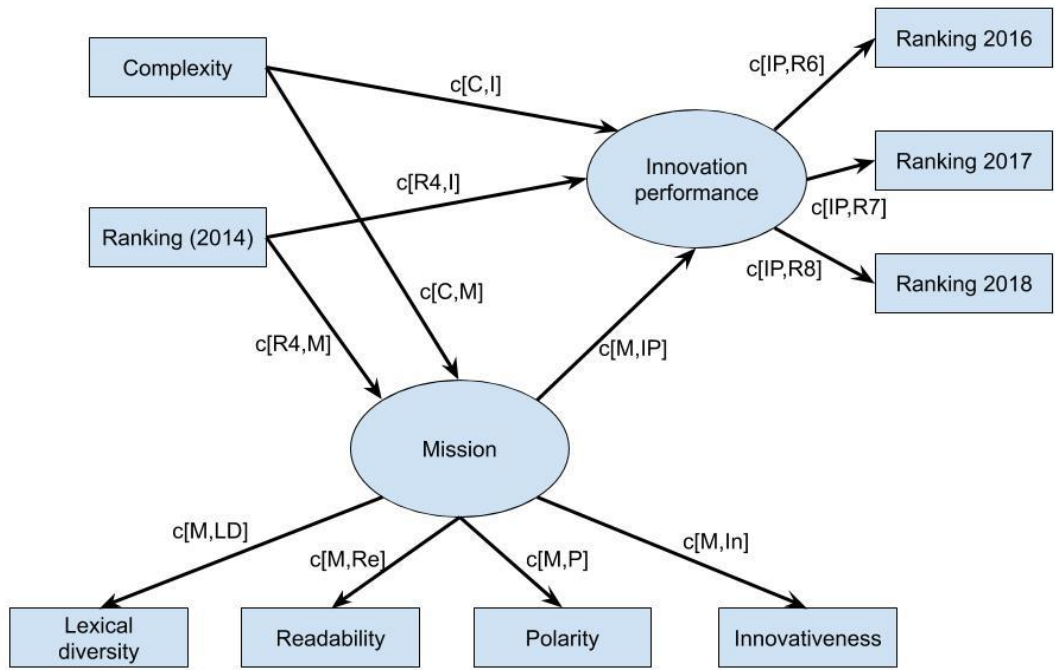

**Figure 3 Structural Equation Modeling Source: the authors' work, based on SCImago (2019) and organizations' websites, processed with R (R Core Team, 2014) package Quanteda (Benoit et al., 2018) for content analysis and the AMOS module for SPSS for SEM (Byrne, 2016; IBM Corp., 2017).**